# Mechanical behaviour of the intact brain-skull interface under tension, compression and shear

Sajjad Arzemanzadeh[a], Karol Miller[a], Ron Kikinis[b,c], Adam Wittek[a, *]

[a] Intelligent Systems for Medicine Laboratory (ISML), School of Engineering, The University of Western Australia, Perth, WA 6009, Australia
[b] Harvard Medical School, Boston, MA, USA
[c] Surgical Planning Laboratory, Department of Radiology, Brigham and Women's Hospital, Boston, MA, USA

* Corresponding author: Adam Wittek (adam.wittek@uwa.edu.au)

## Abstract

The brain-skull interface (meninges) governs the boundary/contact conditions of the brain in computational head models, yet its mechanical behaviour has rarely been measured. We characterised the brain-skull interface by testing brain-skull complex samples (skull bone, meninges and brain tissue in their natural arrangement) excised from 19 sheep cadaver heads under uniaxial tension and compression at strain rates of 0.002, 0.02 and 0.2 $s^{-1}$ and under simple shear at 0.02 $s^{-1}$. Histology confirmed that the meningeal tissue layers remained intact and attached to the brain and skull after the sample excision. We acquired a magnetic resonance image (MRI) of every sample, reconstructed its geometry and built a sample-specific finite element model of each experiment. We represented the mechanical properties of the brain tissue with a second-order Ogden hyperelastic model and the brain-skull interface with a cohesive zone model describing its traction-separation behaviour up to failure. Using inverse finite element analysis, we first calibrated the brain tissue parameters for each subject and then, with these fixed, calibrated the interface parameters against the brain-skull complex sample from the same subject. The shear modulus of the brain tissue increased with strain rate from 0.94 ± 0.16 kPa at 0.002 $s^{-1}$ to 1.60 ± 0.30 kPa at 0.2 $s^{-1}$, whereas the interface parameters (maximum normal traction 2.35 ± 0.81 kPa, maximum tangential traction 4.14 ± 0.97 kPa, fracture energy 0.35 ± 0.14 N m$^{-1}$) showed no statistically significant dependence on strain rate within the range of 0.002-0.2 s$^{-1}$ and no significant difference between shear and tension-compression loading. In every sample, interface failure occurred within the subarachnoid space. Neither a rigid connection nor a frictionless sliding contact with separation, approaches for modelling interactions between the brain and skull commonly used in computational biomechanics head models, reproduced the experimentally measured responses or the failure mode of the interface. Our results provide a comprehensive, experimentally determined cohesive zone model of the brain-skull interface for computational head models.

**Statement of Significance**

Computational predictions for neurosurgical planning and brain injury analysis depend strongly on how the brain-skull connection is represented, yet its mechanical behaviour remains poorly described. We characterised the brain-skull interface through tension, compression, and shear experiments performed at multiple strain rates and demonstrated that a cohesive zone model accurately describes its response. We identified cohesive parameters by inverse finite element analysis of the experiments using sample geometries reconstructed from radiographic images, rather than the idealised geometries commonly assumed in the literature. The identified parameters appeared to be independent of both loading mode and strain rate. Our findings establish a cohesive zone modelling approach for the brain-skull interface and provide constitutive parameters suitable for implementation in comprehensive biomechanical head models.

## 1. Introduction

The brain is anchored within the skull by the brain-skull interface (meninges), the layer of membranous tissue between the cortical surface and the cranium [1, 2]. This layer determines how firmly the brain is held and how it moves when the head is subjected to transient mechanical loading [3, 4]. Its failure is associated with some of the most serious consequences of head trauma, including acute subdural haematoma [5, 6]. The interface comprises three main sublayers: the dura mater, the arachnoid and the pia mater (Figure 1a) [1]. The dura mater is the outermost, consists of dense fibrous connective tissue, and is firmly attached to the inner surface of the skull. The pia mater is the innermost and adheres directly to the cortical surface of the brain. Between the two lies the subarachnoid space, a complex web-like structure that contains the cerebrospinal fluid (CSF) and is traversed by arachnoid trabeculae extending from the arachnoid to the pia mater. In humans, the entire interface is approximately one millimetre thick [2]. Because the three sublayers differ markedly in composition and stiffness, the interface behaves as a composite that may fail within itself rather than pulling away from the bone or from the brain.

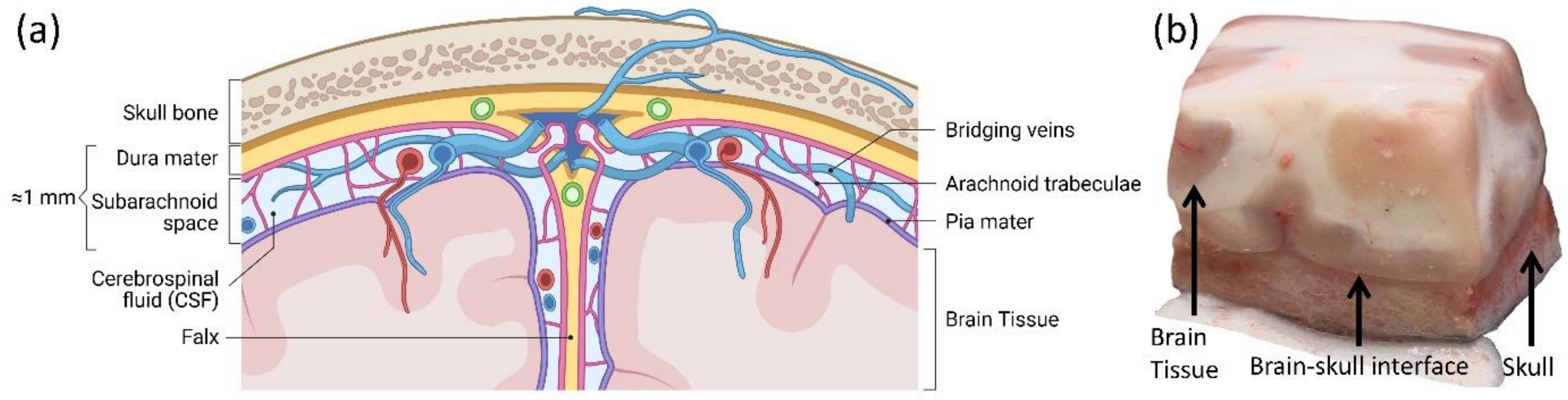

*Figure 1 (a) Schematic of the brain-skull interface (meninges). For humans, the average thickness of the meninges is approximately 1 mm. (b) Brain-skull complex sample excised from a sheep cadaver head in this study, showing the brain tissue, the brain-skull interface and the skull bone.*

The motivation for this study is that results obtained using comprehensive computational biomechanics head models, used in applications ranging from surgical training and planning to traumatic brain injury prediction and prevention, critically depend on specification of boundary and/or contact conditions for the brain that are governed by the brain-skull interface [3, 4, 7, 8]. In the absence of measurements made on the intact interface, the connection between the brain and the skull is instead represented using arbitrary approaches that include rigid attachment of the brain tissue to the skull bone, a frictionless sliding contact between the brain cortical surface and skull, or a compliant solid layer whose properties are assigned rather than determined experimentally [4, 9, 10]. The predicted deformation of the brain and the predicted risk of injury depend crucially on how the brain-skull interface is modelled [4, 11, 12]. To ensure biofidelity of such models, the mechanical behaviour of the brain-skull interface needs to be measured and its mathematical description be determined.

Determining the properties of the brain-skull interface directly is challenging because it is a thin and delicate multi-layer soft tissue composite that is difficult to isolate without damage. Consequently, most of the quantitative data now available were obtained from meningeal layers that form the interface tested after separation from the adjacent tissues [2]. Jin et al. extracted pia-arachnoid complex samples from bovine cadaver heads and loaded them in in-plane tension, normal traction and shear at strain rates between 0.05 and 100 $s^{-1}$, reporting elastic moduli of 6 to 40 MPa in in-plane tension, 61 to 148 kPa under normal traction and 11 to 22 kPa in shear, a difference of three orders of magnitude, which reflects how strongly the measured stiffness depends on the loading mode [13]. Walsh et al. characterised the human dura mater, falx cerebri and superior sagittal sinus under biaxial tension [14]. Such experimental studies of separate meningeal layers are valuable. However, these studies provide only limited information about the behaviour of the intact brain-skull interface, which comprises an assembly of such layers. In particular, they do not provide information on the interface strength and its failure behaviour.

Therefore, in our previous studies, we proposed a dissection protocol in which the brain-skull interface is harvested together with the adjacent skull bone and brain tissue (Figure 1b) from cadaveric heads (ovine heads were used in this study) to minimise the risk of damage to the interface [15, 16]. Preliminary experiments performed on a limited number of the resulting brain-skull complex specimens confirmed both the feasibility and suitability of the proposed protocol. The results further demonstrated that the brain-skull interface can be represented by a cohesive zone model, a mathematical description that relates the traction transmitted across an interface to the separation of its surfaces up to complete failure, which reproduced both the force-displacement behaviour of the intact interface and the onset of its failure [15].

In this study, we applied the dissection protocol and expertise from our previous investigations [15, 16] to comprehensively characterise the mechanical behaviour of the brain-skull interface under tension, compression, and shear, including its strain-rate dependence and failure mechanisms. We performed histological analysis to verify that the meninges remained intact throughout the dissection process. We reconstructed the three-dimensional geometry of each sample from its own magnetic resonance image (MRI) and used it to generate a geometrically accurate computational grid (finite element mesh). Unlike most previous studies, which identify a single set of constitutive parameters from averaged experimental curves, we calibrated the parameters separately for each sample, using the geometry and the brain tissue properties of that sample's subject. We first conducted inverse finite element (FE) analysis to calibrate the subject-specific constitutive parameters of the brain tissue using a second-order Ogden model. Subsequently, with the calibrated brain tissue properties kept constant, we identified the parameters of a cohesive zone model representing the brain-skull interface through inverse FE analysis. We then compared the capability of the proposed cohesive-layer formulation to reproduce the experimentally observed force-displacement responses by comparison with that of two interface modelling approaches commonly used in computational biomechanics head models: 1) Rigid attachment between the brain and skull; and 2) Frictionless sliding contact between the brain and skull [4, 9].

## 2. Methods

Figure 2a summarises the procedure followed for every sheep head. We cut two types of samples from each subject: brain tissue alone and the brain-skull complex. Samples were imaged by MRI, and then mechanical tests were conducted. A sample-specific FE model of each experiment was built from the geometry reconstructed from the MRIs. We then adjusted the material parameters of each model by inverse analysis until the computed force-displacement response matched the measured one. We analysed the two samples in sequence: we first calibrated constitutive parameters of the brain tissue sample and then held the resulting brain tissue parameters fixed in the model of the brain-skull complex sample from the same subject, leaving the parameters of the brain-skull interface as the only unknowns to be determined.

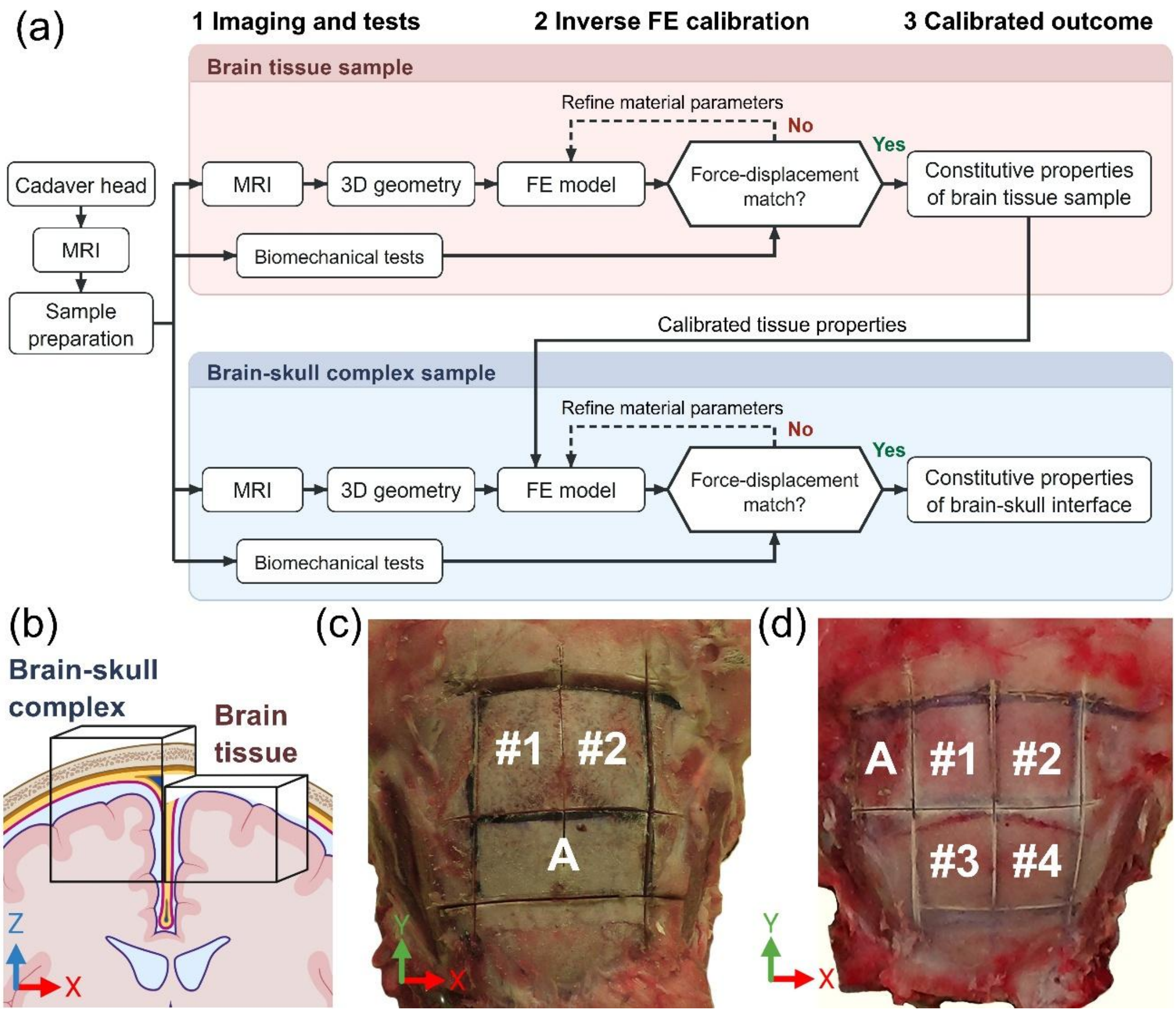

*Figure 2 (a) Workflow we used to calibrate mechanical properties of the intact brain-skull interface through experiments and FE analysis. (b) Schematic of the head showing the regions where the brain-skull complex samples and by the brain tissue samples were extracted. (c, d) Sheep cadaver heads before sample extraction with indicated cutting lines. Region A was removed first to provide access for the horizontal cut beneath the numbered regions, from which the brain-skull complex and brain tissue samples were then excised. In (c), two indicated numbers represent one brain tissue and one brain-skull complex sample used in simple shear tests. In (d), four indicated numbers represent two brain tissue samples and two brain-skull complex samples used in uniaxial normal tests. Samples 1 and 2 have been used for tensile tests, and samples 3 and 4 have been used for compressive tests. The dissection procedure has been designed to extract cuboidal samples with dimensions of 25 mm x 25 mm x 15 mm, for the brain tissue, and 25 mm x 25 mm x 20 mm for the brain-skull complex sample.*

### 2.1. Sample Preparation

Skinned sheep cadaver heads were obtained from a commercial abattoir (Dardanup Butchering Company, Bunbury, WA, Australia) and transported to the Harry Perkins Institute of Medical Research (Nedlands, WA, Australia), where the sample preparation, imaging and mechanical testing were carried out. The heads were kept refrigerated, and every experiment was completed within 18 hours of the animal's death and within 4-6 hours after the sample extraction, which is shorter than post-mortem delays known to significantly alter the mechanical characteristics of the brain tissue [17, 18].

MRI of each head was acquired to verify that the skull and the intracranial contents were not mechanically damaged. We harvested two types of samples from each head: (i) brain tissue samples (containing only the brain tissue) and (ii) brain-skull complex samples consisting of the skull bone, the meninges and the brain tissue in their original arrangement, so that the brain-skull interface remained intact. From each sheep cadaver head (subject) used for tension-compression tests, we extracted four samples: one brain tissue sample and one brain-skull complex sample for the tensile test, and one of each for the compressive test (Figure 2d). The heads used for shear tests were smaller and could not accommodate four samples, so from each of them we harvested two samples, one brain tissue sample and one brain-skull complex sample (Figure 2c). The dissecting tools and the order of the dissection follow our previous studies [15, 16], and the region from which the samples were harvested is explained in Figure 2b-d. Every sample was harvested from a single hemisphere, with one of the dissecting lines on the sagittal suture. This ensured that the samples did not include the falx and the bridging veins, c structures that anchor the brain to the skull far more firmly than the interface and would otherwise dominate the measured responses.

After excision, the bottom surface of the brain tissue samples was immediately attached to custom 3D-printed platens using a fast-curing cyanoacrylate glue (Super Glue by Selleys) to facilitate sample handling without direct contact with the tissue and minimise the risk of inducing damage. For the brain-skull complex samples, the curved outer surface of the skull was built up with an epoxy putty (Knead It Aqua by Selleys). This preparation allowed the top surface of the sample and the brain-skull interface plane to be oriented normal to the tensile and compressive loading directions and aligned with the direction of shear loading.

The samples were kept hydrated during handling and transport within the building, including transfer between the laboratory where dissection was performed, the MRI scanner, and the laboratory where biomechanical testing was conducted. Hydration was maintained by exposing the samples to physiological saline vapour. To achieve such exposure, the samples were placed in a sealed container containing approximately 10 mL of physiological saline solution, without direct contact with the liquid. In total, we prepared and tested 33 brain tissue samples and 33 brain-skull complex samples from 19 subjects: 14 brain tissue and 14 brain-skull complex samples under tensile loading, 14 of each under compressive loading and 5 of each under shear loading (Table 1).

Because the MRI resolution is insufficient to distinguish different meningeal layers and visualise their structure, histological examination was conducted to confirm that our sample excision procedure did not damage the interface tissues. The analysis was carried out at the CELLCentral Histology laboratory of the University of Western Australia. We harvested samples for histological examination from an additional sheep cadaver head that was not used for mechanical testing, from the same region and with the same dissection protocol as the brain-skull complex samples used for biomechanical testing (Figure 2). The specimens were fixed in formalin for five weeks. As cutting/sectioning through the skull bone would require application of relatively large forces that could damage the interface tissues, the specimens were subsequently immersed in an ethylenediaminetetraacetic acid (EDTA) solution for six weeks to decalcify the bone. To enable examination of the meningeal structures, following decalcification, the specimens were sectioned at a nominal thickness of 5 μm and stained with haematoxylin and eosin (H&E), a commonly used histological stain.

### *2.2. Acquisition of Sample Images*

Brain tissue is extremely compliant and adherent; it is therefore very challenging to cut samples that conform to a nominal geometry, typically a cuboidal [19-21] or cylindrical [22-24]. We have previously shown that assuming that samples conform to the nominal geometry when applying inverse analysis to determine the material properties of very soft tissues can lead to an underestimation of the shear modulus by about 10% under shear loading and nearly 50% under axial compression [19]. The issue is even more significant for brain-skull complex samples because of the irregularly curved skull surface. Therefore, to obtain information about the sample geometry, MRIs of all samples were acquired before the biomechanical testing. Samples for the tension and compression tests were scanned on a Bruker Biospec 9.4 Tesla animal MRI scanner using a three-dimensional T2-weighted TurboRARE sequence (repetition time 2200 ms, echo time 34 ms, RARE factor 12, matrix size 396 x 192 x 48), giving a voxel size of 0.25 mm x 0.25 mm x 0.5 mm. Samples for the shear tests were scanned on a Siemens Healthineers Vida 3 Tesla scanner using a two-dimensional T2-weighted turbo spin echo sequence (repetition time 15830 ms, echo time 93 ms, 11 averages, 64 slices, slice thickness 0.6 mm, turbo factor 18, matrix size 320 x 220), giving isotropic 0.6 mm voxels. The scanner change was necessitated by a failure of the 9.4 T Bruker

BioSpec scanner during the study. As T2 weighting is a critical parameter in tissue segmentation, the sequence implemented on the replacement Siemens Healthineers scanner was selected to preserve the T2 weighting used on the Bruker BioSpec scanner.

### *2.3. Biomechanical Testing*

Figure 3 shows the experimental setups for biomechanical testing. The experiments were performed on a UniVert biomaterial testing system (CellScale) at the T3mPLATE Laboratory of the Harry Perkins Institute of Medical Research (Nedlands, Western Australia, Australia). Information about the loading modes, loading speeds and load cells used is given in Table 1. The experimental protocol follows the general guidelines of Miller [25, 26] and Miller and Chinzei [23, 24]. The 3D-printed fixtures for handling the samples were rigidly attached to the base plate of the testing machine. No-slip boundary conditions between the sample top surface and the UniVert testing system loading platen were created. In compressive tests, this was achieved by attaching an 80-grit sandpaper to the loading platen. In tensile and shear tests, the sample top surface was glued to the loading platen using fast-curing cyanoacrylate glue. To ensure adhesion, after the initial contact of the sample top surface with the loading platen, the sample was compressed by 1 mm for 30 s. It was then returned to the original unloaded configuration and held there for an additional 180 s to provide stress relaxation before starting the tests. The time of 180 s was selected since it has been reported in the experimental literature as sufficient for brain tissue relaxation [27]. Tension and compression were applied at the loading platen speeds of 0.03, 0.3 and 3 mm/s, corresponding to nominal strain rates of approximately 0.002, 0.02 and 0.2 $s^{-1}$, respectively. For shear, the loading platen speed was 0.3 mm/s, which corresponds to a nominal shear strain rate of approximately 0.02 $s^{-1}$. All tests were conducted at room temperature (20-22 °C). Three digital cameras recorded each experiment from the front and from two oblique directions, and these recordings were examined frame by frame to establish when the interface began to fail and how the failure propagated.

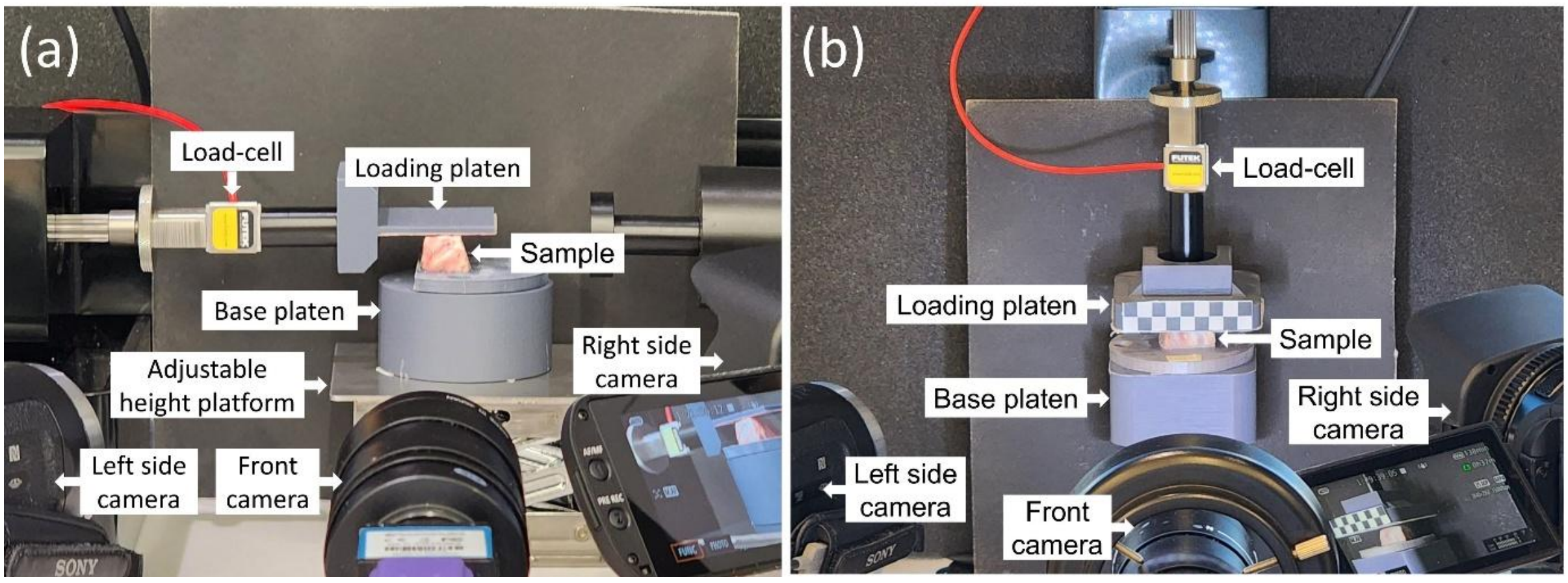

*Figure 3 Experimental setup used in this study. (a) Setup used for the tests using simple shear. (b) Setup used for the tests using uniaxial tension and compression.*

*Table 1 Test matrix used in this study. Samples extracted from each sheep cadaver head were tested only under one set of the experimental conditions (referred to as the loading groups) defined by the loading mode and loading speed (and nominal strain rate). Each sample was tested only once. For uniaxial tension and compression, four samples were extracted from each head: one brain tissue sample and one brain-skull complex sample tested under tensile loading, and one of each tested under compressive loading. The tensile and compressive tests for the samples sourced from a given head subject were analysed together, so that one set of brain tissue parameters and one set of brain-skull interface parameters were calibrated simultaneously against both tests (Section 2.4.4). For simple shear tests, two samples were extracted from each head: one brain tissue sample and one brain-skull complex sample. The nominal dimensions of the brain tissue samples were 25 mm x 25 mm x 15 mm, and the brain-skull complex samples were 25 mm x 25 mm x 20 mm.*

| Loading mode | Loading speed (mm $s^{-1}$) | Nominal strain rate ($s^{-1}$) | Loading Group | Number of subjects (cadaver heads) | Number of samples per subject | Load cell (accuracy) |
|---|---|---|---|---|---|---|
| Uniaxial Tension and compression | 0.03 | 0.002 | 1 | 5 | 4 | 10 N (0.2%, 20 mN) |
| | 0.3 | 0.02 | 2 | 5 | 4 | 2.5 N (0.1%, 2.5 mN) |
| | 3 | 0.2 | 3 | 4 | 4 | |
| Simple shear | 0.3 | 0.02 | 4 | 5 | 2 | |

### *2.4. Finite Element Models of the Experiments*

#### *2.4.1 Geometry Reconstruction and Mesh Generation*

The process of constructing FE meshes (computational grids) is shown in Figure 4. We segmented MRIs of the brain tissue samples semi-automatically in 3D Slicer, a freely available open-source software platform for image analysis and 3D visualisation [28], using intensity thresholding followed by Gaussian smoothing tools within the 3D Slicer Segment Editor module [29], which is an adequate approach when the sample can be regarded as consisting of a single soft material surrounded by air. The brain-skull complex samples, however, consist of different materials that cannot be reliably delineated using thresholding alone. Therefore, for the MRIs of brain-skull complex samples, we used a deep learning segmentation model implemented in the MONAI framework [30, 31]. We trained the model on MRIs of eight brain-skull complex samples that we had segmented manually, slice by slice, into brain tissue and skull bone. The trained model was then applied to the remaining samples to segment them automatically. We inspected each resultant segmentation visually, slice by slice, and compared it against its MRI. We created computational grids (FE meshes) using the Coreform Cubit finite element mesh generator [32], widely used for hexahedral meshes, from the 3D surfaces reconstructed from the segmented MRIs of the brain tissue and brain-skull interface samples. The element size was 0.5 mm, which matches the resolution of the images and ensures an h-convergence (mesh independence) of the solution obtained using the finite element method. For all samples, the mean scaled Jacobian of the brain tissue mesh elements exceeded 0.95, indicating high mesh quality according to criteria reported in the literature [3].

### *2.4.2. Finite Element Formulation, Boundary Conditions and Loading*

All analyses were conducted using the non-linear explicit dynamics FE procedures from the Abaqus 2023 finite element code (Dassault Systèmes, Providence, RI, USA) [33] with automatic time stepping to ensure solution stability and accuracy. The initial time increment was set to $2.5\times10^{-5}$ s in all simulations. We used 8-noded under-integrated hexahedral elements (C3D8R element type in Abaqus finite element code) with stiffness-based hourglass control. This element does not exhibit volumetric locking, which makes it appropriate for modelling of the brain tissue and other nearly incompressible continua.

The brain-skull interface was represented by a single layer of 8-noded hexahedral cohesive elements (element type COH3D8 in Abaqus FE code) inserted between the brain tissue and the skull. Boundary conditions were defined to replicate the experimental setup. In the FE models of brain tissue samples, the bottom nodes of the samples were rigidly constrained as in the experiments, the bottom (dorsal) surfaces of the samples were glued to the base platen. In the FE models of brain-skull complex samples, the skull bones were rigidly constrained as in the experiments; the outer surface of the skull bones was glued to the base platen. For tensile and shear tests, the loading was applied to the sample top surface nodes via a prescribed displacement. Nodes for prescribing the displacements were identified from video recordings of the tests. For compressive tests, we modelled the loading and base platen as rigid plates and used the no-slip contact algorithm from the Abaqus finite element code between the sample top surface and the moving loading platen, to replicate the no-slip condition provided by the sandpaper attached to the loading platen in the experiments. The displacements were defined using a smooth step procedure that uses a fifth-order polynomial function in the Abaqus FE software. A typical sample-specific simulation took approximately 10 min for brain tissue samples and 20 min for brain-skull complex samples on 8 cores of an Intel Core i9-13900K CPU.

### *2.4.3. Material Models*

We modelled the skull bone as rigid since it is orders of magnitude stiffer than the brain tissue. Each subject was tested at one of three constant loading speeds (Table 1). We did not aim to develop a time-dependent (viscoelastic) model of the brain tissue; we needed only the brain tissue properties of each subject under the same loading conditions as its brain-skull complex sample, as input for calibrating the brain-skull interface properties of that sample. We therefore described the brain tissue behaviour by a second-order Ogden hyperelastic strain energy function [34] with parameters identified separately for each subject at its tested strain rate:

$$W\left(\overline{\lambda_1}, \overline{\lambda_2}, \overline{\lambda_3}\right) = \sum_{i=1}^{2} \frac{2\mu_i}{\alpha_i^2}\left(\overline{\lambda_1}^{\alpha_i} + \overline{\lambda_2}^{\alpha_i} + \overline{\lambda_3}^{\alpha_i} - 3\right) + \sum_{i=1}^{2} \frac{1}{D_i}(J_{el} - 1)^{2i}, \qquad (1)$$

in which $\overline{\lambda_i}$ are deviatoric principal stretches, $J_{el}$ is the elastic volume change, $\mu_i$ are the shear parameters, $\alpha_i$ are the dimensionless constants, and $D_i$ are material constants determining compressibility. The shear modulus is given by $\mu = \mu_1 + \mu_2$. As the brain is nearly incompressible, we used a Poisson's ratio of 0.49. Application of the Ogden hyperelastic strain energy function for the brain tissue is a well-established practice [20, 26, 35].

In our experiments, the brain-skull interface remained adherent to both the brain tissue and the skull bone, transmitting increasing mechanical load until it exhibited softening and subsequent failure confined to a thin layer within the interface (Section 3.3). This type of behaviour has been described in the literature using cohesive zone models [36]. Such models are implemented in major finite element packages and enable modelling of thin composite structures composed of multiple materials undergoing failure using a single layer of cohesive elements [33, 37]. We therefore discretised the brain-skull interface using a single layer of hexahedral elements described by a traction-separation cohesive zone model (CZM) defined as:

$$\begin{Bmatrix} t_n \\ t_{s_1} \\ t_{s_2} \end{Bmatrix} = \begin{pmatrix} E_n & 0 & 0 \\ 0 & E_{s_1} & 0 \\ 0 & 0 & E_{s_2} \end{pmatrix} \begin{Bmatrix} \varepsilon_n \\ \varepsilon_{s_1} \\ \varepsilon_{s_2} \end{Bmatrix}, \tag{2}$$

where $E_n$, $E_{s_1}$, and $E_{s_2}$ are linear elastic traction moduli, $t_n$ is nominal traction in the normal direction, $t_{s_1}$ and $t_{s_2}$ are the nominal tractions in the two local shear directions, $\varepsilon_n$ is the nominal strain in the normal direction, $\varepsilon_{s_1}$ and $\varepsilon_{s_2}$ are the nominal strains in the two local shear directions. We selected the quadratic nominal stress ratio as the damage initiation criterion as it couples the normal and tangential terms, so both influence the computed response in every loading configuration:

$$\left\{\frac{\langle t_n \rangle}{t_n^0}\right\}^2 + \left\{\frac{t_{s_1}}{t_{s_1}^0}\right\}^2 + \left\{\frac{t_{s_2}}{t_{s_2}^0}\right\}^2 = 1, \tag{3}$$

where $t_n^0$ denotes the maximum normal traction, $t_{s_1}^0$ and $t_{s_2}^0$ are the maximum tangential tractions, and the Macaulay bracket of normal traction, $\langle t_n \rangle$, prevents a compressive normal traction from contributing to damage. Brain-skull interface sublayers such as the dura mater and the pia-arachnoid complex have been reported to be transversely isotropic [2, 13, 14, 38, 39], which is consistent with our histological analysis of the brain-skull interface, showing no obvious structural difference between the two tangential directions in the regions examined (Figure 7). We therefore assumed isotropy in the tangential plane, i.e. $E_{s_1}$=$E_{s_2}$ and $t_{s_1}^0 = t_{s_2}^0$, referred to hereafter as $E_s$ and $t_s^0$ in the following sections, respectively.

The damage evolution was modelled with an energy-based evolution law with linear softening, characterised by the fracture energy G. The elastic tractions determined by moduli $E_n$ and $E_s$, govern the brain-skull interface response only up to the onset of damage. As the brain-skull interface acts in series with the brain tissue and is substantially stiffer than the brain tissue, its elastic moduli cannot be determined reliably from the force-displacement response of composite specimens comprising both the brain tissue and the brain-skull interface. Therefore, we obtained these properties from the experiments on pia-arachnoid complex specimens reported by Jin et al., who measured the normal $E_n = 61\ kPa$ and tangential $E_s = 11\ kPa$ moduli at strain rates comparable to those used in the present study [13, 38]. Consequently, the maximum tractions ($t_n^0$ and $t_s^0$) and fracture energy ($G$) were the parameters of the cohesive zone model of the brain-skull interface that needed to be calibrated to match experimental force-displacement results of the brain-skull complex samples under loading.

To compare the performance of our approach using the cohesive zone model with the methods commonly used in the literature [4, 9], models of the experiments in which the brain-skull complex samples were subjected to tension and compression (samples from subject H6; strain rate of 0.02 $s^{-1}$, see Table 2) and shear (sample from H15, see Table 2) were analysed also with the brain-skull interface represented by: (i) a rigid connection between brain tissue and the skull bone, and (ii) a frictionless sliding contact with separation between the brain tissue and the skull.

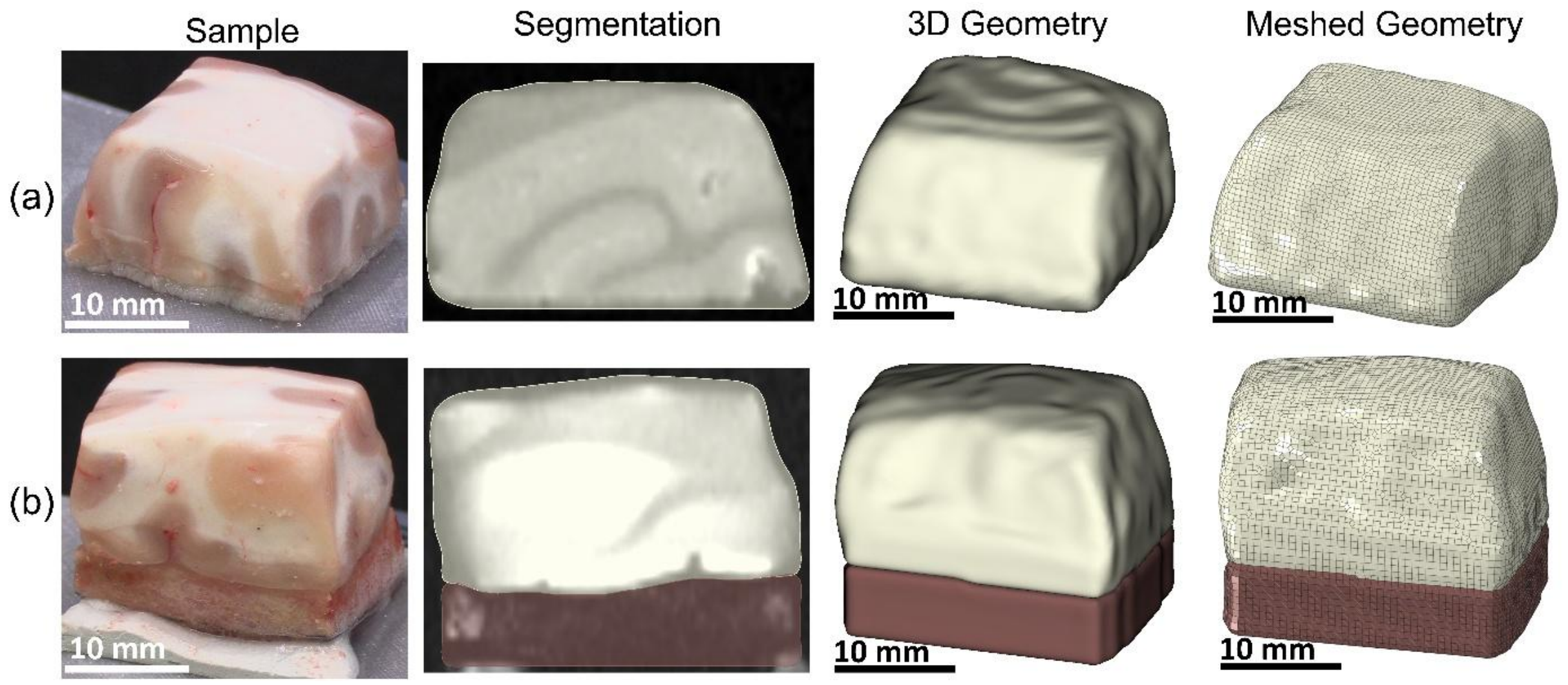

*Figure 4 Construction of the FE models of: (a) a typical brain tissue sample and (b) a typical brain-skull complex sample used in this study. First column: experimental samples, second column: the MRI of the samples with highlighted material segmentation, third column: 3D representation of the segmentation, and fourth column: meshed geometries imported into Abaqus. White colour: brain tissue; Red colour: skull bone. The mesh in (a) has 40633 elements. The mesh in (b) has 38369 elements in the brain tissue, 1719 cohesive elements in brain-skull interface and 19148 elements in the skull bone.*

### *2.4.4. Determination of Material Parameters*

We determined the material constants of the brain tissue and the brain-skull interface in two stages, separately for each subject and therefore for the strain rate at which that subject was tested. First, we calibrated the subject-specific constitutive parameters ($\mu_1$, $\mu_2$, $\alpha_1$ and $\alpha_2$) of the brain tissue using the results of the experiments conducted on the brain tissue samples of that subject. These parameters were then used to describe the brain tissue within the model of the brain-skull complex sample from the same subject and at a given strain rate, leaving $t_n^0$, $t_s^0 = t_t^0$ and G as the only parameters to be calibrated. The calibration was formulated as an optimisation problem in which the parameters are optimised so that the force-displacement characteristics obtained from computer simulation of the biomechanical tests of either brain tissue samples or the brain-skull complex samples match those measured in the tests. The calibration was conducted using our in-house inverse FE software with the optimisation problem solved using the established Optuna framework [40] and its covariance matrix adaptation evolution strategy (CMA-ES) sampler [40, 41]. Parameters of the brain tissue under tensile and compressive loading were calibrated simultaneously against the results of compressive and tensile tests conducted using the brain tissue samples from the same subject, with the objective function:

$$cost_{Brain} = 0.5\, NRMSE_{Comp} + 0.5\, NRMSE_{Ten}, \tag{4}$$

where $NRMSE_{Comp}$ and $NRMSE_{Ten}$ are the root-mean-square force residuals of the compressive and tensile branches, each normalised by the peak force of that branch. When calibrating the constitutive parameters of the brain-skull interface, the objective function was:

$$cost_{Interface} = 0.25\, NRMSE_{Comp} + 0.25\, NRMSE_{Ten} + 0.25 \left|\frac{\Delta F_f}{F_{f,exp}}\right| + 0.25 \left|\frac{\Delta D_f}{D_{f,exp}}\right|, \tag{5}$$

where $F_{f,exp}$ and $D_{f,exp}$ are the force and displacement at interface failure measured in the experiments, and $\Delta F_f$ and $\Delta D_f$ are the differences between the experimental and FE-predicted values at the failure. The terms $\left|\frac{\Delta F_f}{F_{f,exp}}\right|$ and $\left|\frac{\Delta D_f}{D_{f,exp}}\right|$ therefore penalise discrepancies in the failure point, which was determined from the force drop together with inspection of the video recordings. For shear tests, both $NRMSE_{Comp}$ and $NRMSE_{Ten}$ terms in Eqs. (4) and (5) were replaced with $NRMSE_{shear}$. For most analysed brain-skull interface samples, the failure occurred at the maximum of the tensile/shear force. The $NRMSE$ term measures how well the whole curve is reproduced and the penalty terms how well the failure event is represented. We ran at least 100 optimisation trials per sample, with the set of parameters obtained from the CMA-ES sampler.

### 2.5. Strain rate dependency and comparison between shear and normal axial loading

Based on Table 1, we have four loading groups: (i) tension and compression at a strain rate of 0.002 $s^{-1}$ (five subjects), (ii) tension and compression at 0.02 $s^{-1}$ (five subjects), (iii) tension and compression at 0.2 $s^{-1}$ (four subjects), and (iv) simple shear at 0.02 $s^{-1}$ (five subjects). The size of each loading group equals its number of subjects, and each subject contributes a single set of calibrated parameters reported as mean ± standard deviation. To quantitatively assess whether the calibrated parameters of the brain tissue (μ) and of the brain-skull interface ($t_n^0$, $t_s^0$ and G) depend on strain rate under tension-compression loading, and whether they differ between shear and tension-compression loading, we used two non-parametric rank-based tests under the null hypothesis that the loading groups yield the same distribution of the parameter.

We assessed the dependence on strain rate with the Kruskal-Wallis test [42], which compares three or more independent groups and was therefore applied to groups (i), (ii) and (iii). The Kruskal-Wallis test assesses whether the ranks of the pooled observations are spread evenly among the groups or concentrated in some of them.

We assessed the difference between loading modes with the Mann-Whitney U test [43], which compares two independent groups. It was applied to group (ii) and group (iv), so that all subjects had the same strain rate of 0.02 $s^{-1}$. Mann-Whitney U test evaluates whether values from one group tend to be larger than values from the other.

For both Kruskal-Wallis and Mann-Whitney U tests, we computed exact p-values of the null hypothesis rather than the usual large-sample approximations, which are unreliable for group sizes of four or five [44].

## 3. Results

### *3.1. Mechanical Behaviour and Constitutive Parameters of the Brain Tissue*

An initial four-parameter optimisation showed that the exponents remained effectively constant across samples and loading modes, with $\alpha_1$ close to -8 and $\alpha_2$ close to 16, which justified holding them fixed in the two-parameter calibration of $\mu_1$ and $\mu_2$ (Section 2.4.4). Figure 5 shows representative force-displacement curves of the brain tissue samples for each strain rate in tension and compression and in shear. The curves of all 19 subjects are given in Supplementary Figure S1. Our samples exhibited the tension-compression asymmetry characteristic of the brain tissue, with the force at a given displacement several times larger in compression than in tension.

Table 2 reports the calibrated parameters of the brain tissue for every subject. The shear modulus of the brain tissue, $\mu = \mu_1 + \mu_2$, under tension-compression loading was 0.94 ± 0.16 kPa at 0.002 $s^{-1}$, 1.52 ± 0.18 kPa at 0.02 $s^{-1}$ and 1.60 ± 0.30 kPa at 0.2 $s^{-1}$. Our results suggest that the shear modulus of the brain tissue is strain-rate dependent, and this was quantitatively confirmed by conducting exact Kruskal-Wallis test under the null hypothesis that there are no differences between the distributions of the brain tissue shear modulus at the three strain rates analysed in this study which was rejected at $p = 0.0028$ (Table 3).

The shear modulus identified from the shear tests at 0.02 $s^{-1}$ was 1.36 ± 0.36 kPa and tended to be somewhat smaller than that identified from the tension and compression tests (1.52± 0.18 kPa) at the same rate. However, this difference was not significant (exact Mann-Whitney U test, p = 0.69, Table 3). The key difference between the results from shear test and those from the tension and compression tests was the contribution of the $\mu_2$ term (associated with the exponent $\alpha_2$ = 16) to the shear modulus, $\mu = \mu_1 + \mu_2$. In the shear tests, $\mu_2$ accounted for 32% ± 10% of μ, compared with only 14% ± 6% in tension and compression (Table 2). This is consistent with the fact that the stiffness parameter associated with a negative exponent dominates the response in compression and at low tensile/shear strains, whereas the parameter associated with a positive exponent dominates at high tensile and shear strain.

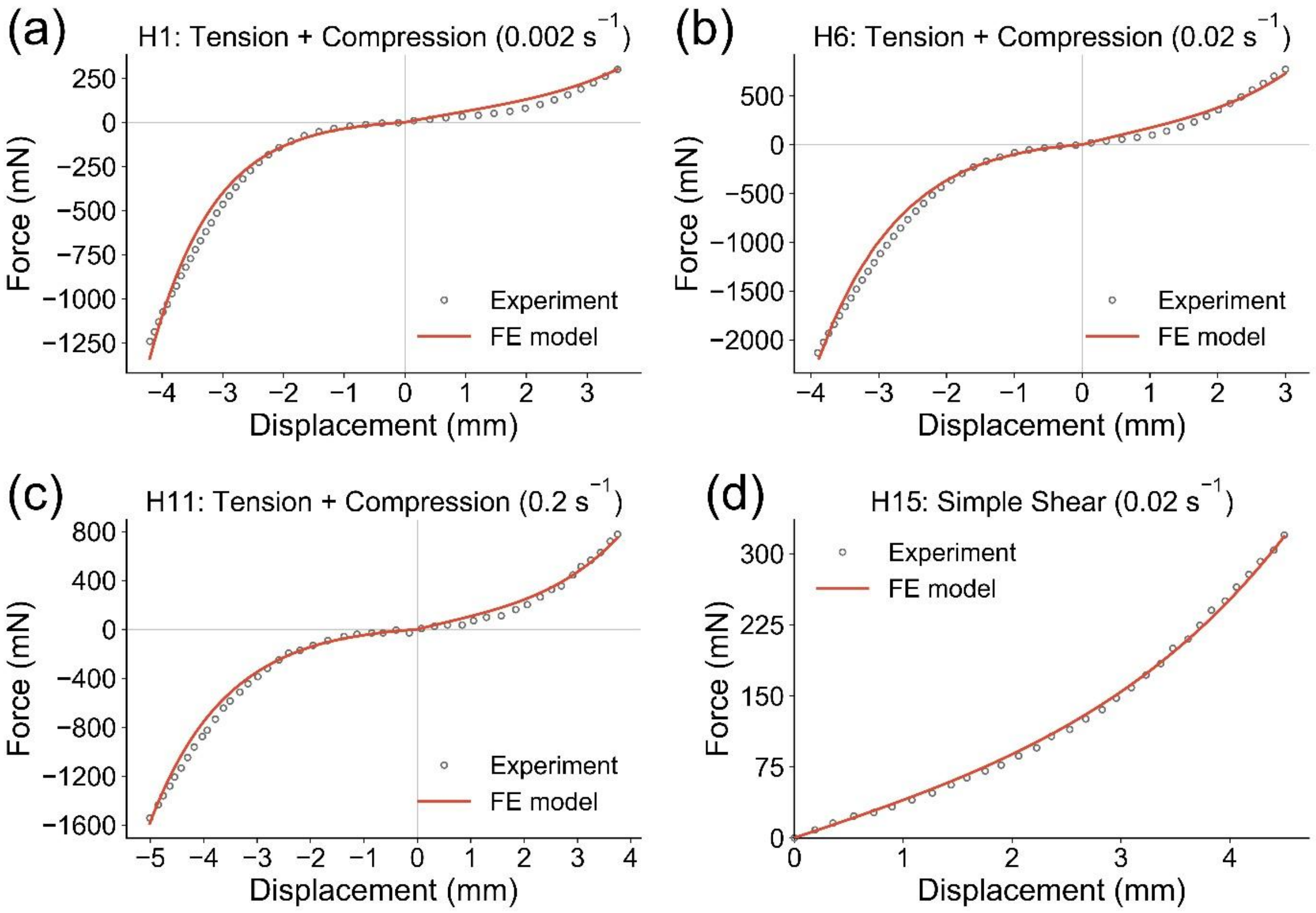

*Figure 5 Typical force-displacement characteristics of brain tissue samples, each from a different subject (H1, H6, H11 and H15; Table 2): experiments (grey hollow circles) and sample-specific FE model of the same experiment (red solid line), with the shear modulus of the brain tissue calibrated for that subject. (a) Subject H1: tension and compression tests at a strain rate of 0.002 $s^{-1}$; (b) Subject H6: tension and compression tests at a strain rate of 0.02 $s^{-1}$; (c) Subject H11: tension and compression tests at a strain rate of 0.2 $s^{-1}$; (d) Subject H15: simple shear*

*test at a shear strain rate of 0.02 s⁻¹. Negative displacement and force denote compression. The responses of all 19 subjects are given in Supplementary Figure S1.*

Table 2 Calibrated constitutive parameters of the brain tissue (second-order Ogden model with $\alpha_1$ = -8 and $\alpha_2$ = 16) and of the brain-skull interface (cohesive zone model) for each subject. $\mu = \mu_1 + \mu_2$ is the shear modulus of the brain tissue, $t_n^0$ is the maximum normal traction, $t_s^0$ is the maximum tangential traction and G is the fracture energy of the interface.

| Loading | Subject | $\mu_1$ (Pa) | $\mu_2$ (Pa) | $\mu$ (Pa) | $t_n^0$ (kPa) | $t_s^0$ (kPa) | G (N m$^{-1}$) |
|---|---|---|---|---|---|---|---|
| Tension + compression (0.002 s$^{-1}$) | H1 | 708 | 82 | 790 | 2.08 | 3.25 | 0.25 |
| | H2 | 829 | 115 | 944 | 1.36 | 4.54 | 0.25 |
| | H3 | 607 | 161 | 768 | 2.50 | 4.22 | 0.54 |
| | H4 | 884 | 219 | 1103 | 2.48 | 2.99 | 0.34 |
| | H5 | 986 | 104 | 1090 | 2.11 | 4.12 | 0.27 |
| | Mean ± SD | 803 ± 149 | 136 ± 54 | 939 ± 159 | 2.10 ± 0.46 | 3.83 ± 0.67 | 0.33 ± 0.12 |
| Tension + compression (0.02 s$^{-1}$) | H6 | 1417 | 216 | 1633 | 2.12 | 4.67 | 0.52 |
| | H7 | 1142 | 115 | 1257 | 2.39 | 3.19 | 0.24 |
| | H8 | 1390 | 344 | 1735 | 1.46 | 4.58 | 0.21 |
| | H9 | 1221 | 290 | 1511 | 1.65 | 6.02 | 0.20 |
| | H10 | 1186 | 274 | 1460 | 2.10 | 4.98 | 0.45 |
| | Mean ± SD | 1271 ± 125 | 248 ± 87 | 1519 ± 182 | 1.94 ± 0.38 | 4.69 ± 1.01 | 0.32 ± 0.15 |
| Tension + compression (0.2 s$^{-1}$) | H11 | 1059 | 287 | 1346 | 3.46 | 5.11 | 0.49 |
| | H12 | 1857 | 111 | 1967 | 1.68 | 4.35 | 0.23 |
| | H13 | 1231 | 139 | 1370 | 1.46 | 4.38 | 0.31 |
| | H14 | 1583 | 149 | 1731 | 4.42 | 4.97 | 0.66 |
| | Mean ± SD | 1432 ± 357 | 171 ± 79 | 1604 ± 300 | 2.75 ± 1.43 | 4.70 ± 0.39 | 0.42 ± 0.19 |
| Simple Shear (0.02 s$^{-1}$) | H15 | 638 | 474 | 1112 | 3.40 | 2.10 | 0.51 |
| | H16 | 1466 | 285 | 1751 | 2.24 | 4.64 | 0.34 |
| | H17 | 912 | 575 | 1487 | 2.05 | 4.36 | 0.25 |
| | H18 | 581 | 286 | 867 | 3.51 | 2.51 | 0.22 |
| | H19 | 1089 | 478 | 1567 | 2.23 | 3.72 | 0.40 |
| | Mean ± SD | 937 ± 360 | 420 ± 129 | 1357 ± 359 | 2.68 ± 0.71 | 3.47 ± 1.12 | 0.35 ± 0.12 |
| All subjects mean ± SD | | | | | 2.35 ± 0.81 | 4.14 ± 0.97 | 0.35 ± 0.14 |

*Table 3 P-values of the statistical tests of the null hypothesis that the calibrated parameters of the brain tissue (second-order Ogden model) and the brain-skull interface (cohesive zone model). $\mu = \mu_1 + \mu_2$ is the shear modulus of the brain tissue, $t_n^0$ is the maximum normal traction, $t_s^0$ is the maximum tangential traction and G is the fracture energy of the cohesive zone model of the brain-skull interface. The exact Kruskal-Wallis test compares the three tension-compression loading groups (strain rates of 0.002, 0.02 and 0.2 $s^{-1}$; five, five and four subjects), and the Mann-Whitney U test compares the shear and tension-compression loading groups at the same strain rate of 0.02 $s^{-1}$ (five subjects each). All p-values are exact (Section 2.5).*

| Statistical test | P-values | | | |
|---|---|---|---|---|
| | $\mu$ | $t_n^0$ | $t_s^0$ | G |
| Strain-rate dependence, tension-compression (exact Kruskal-Wallis: 0.002, 0.02 and 0.2 $s^{-1}$) | 0.0028 | 0.82 | 0.09 | 0.52 |
| Shear vs tension-compression at 0.02 $s^{-1}$ (exact Mann-Whitney U) | 0.69 | 0.15 | 0.10 | 0.69 |

### ***3.2. Mechanical Behaviour and Constitutive Parameters of the Brain-Skull Interface***

Figure *6* shows representative force-displacement curves of the brain-skull complex samples at each strain rate in tension and compression and in shear (All curves are given in Supplementary Figure S2). The calibrated models reproduced the measured curves closely. The penalty terms of the cost function (Eq. 5), $\left|\frac{\Delta F_f}{F_{f,exp}}\right|$ and $\left|\frac{\Delta D_f}{D_{f,exp}}\right|$, which measure the relative differences between the measured and predicted force and displacement at brain-skull interface failure for each subject, averaged 5.1% and 7.0%, respectively, over the 14 tensile and 5 shear tests.

The brain-skull complex samples behaved like the brain tissue samples in compression, where the interface transmits load without separating, but not in tension or shear. In all tension and shear tests, the force dropped after reaching its peak as the interface failed. Failure progressed more gradually in shear than in tension.

Across all 19 subjects, the maximum normal traction ($t_n^0$) was 2.35 ± 0.81 kPa, the maximum tangential traction ($t_s^0$) was 4.14 ± 0.97 kPa and the fracture energy (G) was 0.35 ± 0.14 N m$^{-1}$ (Table 2). In contrast to the shear modulus of the brain tissue, none of the three interface parameters depended significantly on the strain rate in tension-compression loading (exact Kruskal-Wallis test, Table 3). The higher mean values of $t_n^0$ and G at 0.2 s$^{-1}$ are due to results of the experiments using the samples from two of the four subjects (H11 and H14) and are accompanied by correspondingly large standard deviations. The parameters identified from the shear tests (Table 2) did not significantly differ from those identified from the tension and compression tests at the same strain rate of 0.02 s$^{-1}$ (exact Mann-Whitney U test, Table 3).

In the cohesive element formulation used here (Eqs. 2 and 3), the Macaulay bracket excludes compressive normal traction from damage initiation, so under compressive loading damage initiation is governed by the tangential traction. Therefore, $t_s^0$ is the main parameter controlling damage initiation under compressive loading. Consistent with this, we observed no complete interface failure in the compressive tests, and in the FE models, only a limited number of cohesive elements failed, at approximately 30% brain tissue strain.

To assess how well the average calibrated cohesive zone model parameters can represent an individual subject, we repeated the simulations in Figure 6 with the mean brain-skull interface cohesive zone model parameters of all 19 subjects (Table 2), keeping the subject-specific calibrated brain tissue parameters of each subject. Up to the onset of interface damage, the force-displacement curves were close to those obtained with the subject-specific interface parameters. The relative differences between the measured and predicted force and displacement at brain-skull interface failure, $\left|\frac{\Delta F_f}{F_{f,exp}}\right|$ and $\left|\frac{\Delta D_f}{D_{f,exp}}\right|$, ranged from 5% to 20% and from 11% to 27%, respectively, over the four typical experiments, compared with averages of 5.1% and 7.0% over all 19 tests with the subject-specific parameters. The failure point was overestimated for some subjects (H1, H15) and underestimated for others (H6, H11).

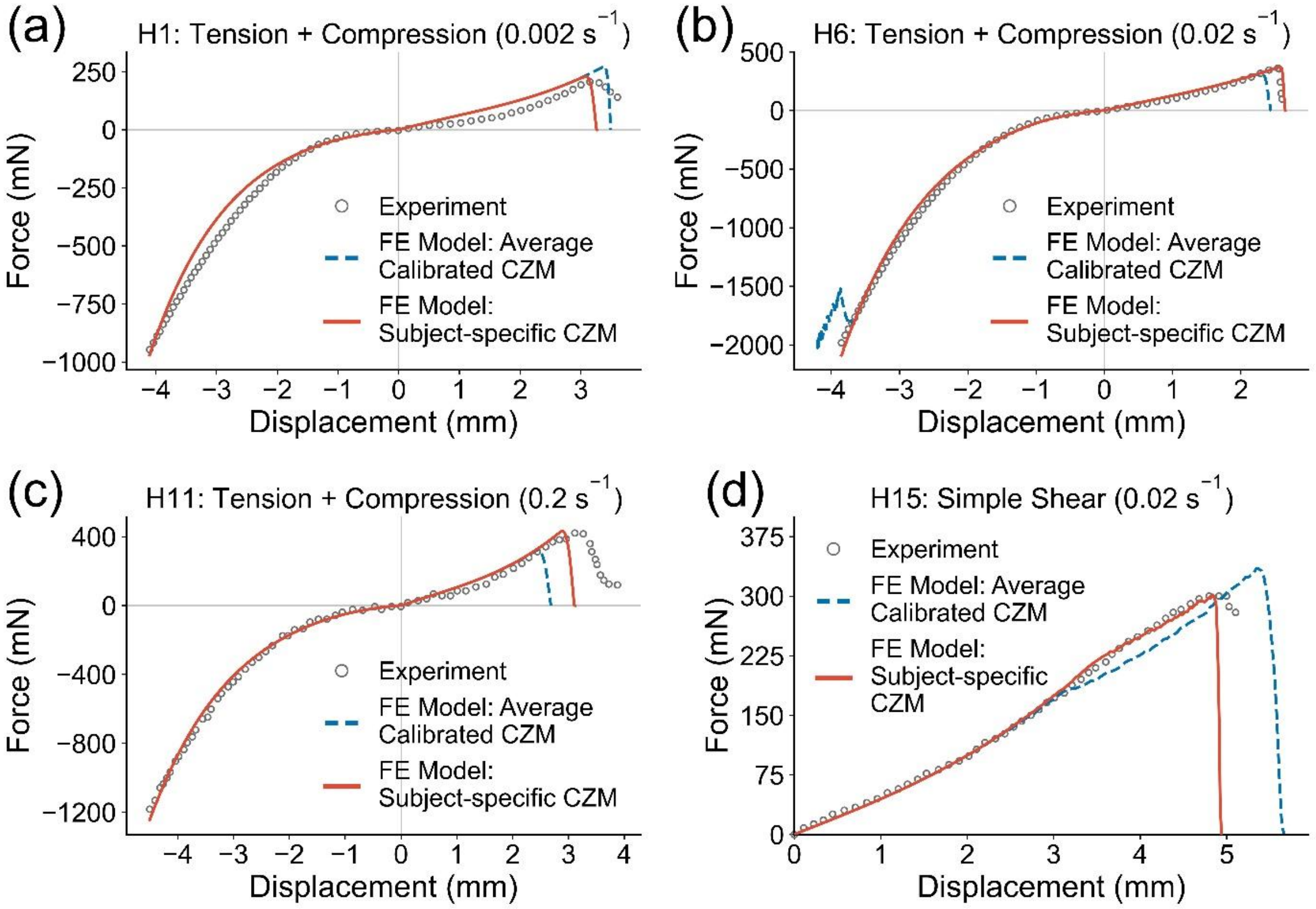

*Figure 6 Typical force-displacement characteristics of brain-skull complex samples from the same subjects as in Figure 5 (H1, H6, H11 and H15; Table 2): experimental results (grey hollow circles) and the sample-specific FE model of the same experiment, with the brain-skull interface represented by a cohesive zone model (CZM) with the brain-skull interface parameters ($t_n^0$, $t_s^0$ and G) calibrated for that subject (red solid lines) or the mean interface parameters of all 19 subjects (blue dashed lines, Table 2). (a) Subject H1: tension and compression tests at a strain rate of 0.002 $s^{-1}$; (b) Subject H6: tension and compression tests at a strain rate of 0.02 $s^{-1}$; (c) Subject H11: tension and compression tests at a strain rate of 0.2 $s^{-1}$; (d) Subject H15: simple shear test at a shear strain rate of 0.02 $s^{-1}$. Negative displacement and force denote compression. The drop in force in tension and in shear marks failure of the brain-skull interface. The responses of all 19 subjects are given in Supplementary Figure S2.*

### *3.3. Validity of Our Protocol for Biomechanical Testing of Brain-Skull Interface: Histological Evidence*

Figure 7 shows a histological image of a brain-skull complex sample, extracted in the same way as the brain-skull complex samples tested in this study. The image indicates that the brain-skull interface remained intact after sample preparation, so any failure in mechanical tests would be associated with the loading applied. Inspection of the brain-skull complex samples after the tests showed that in every sample the separation occurred within the subarachnoid space: the dura mater remained attached to the skull bone, the pia mater remained on the cortical surface of the brain, and the brain tissue itself did not tear. The failure pattern was the same in tension and in shear and at all three strain rates. This is consistent with the histological image (Figure 7), in which the subarachnoid space, bridged only by sparse soft trabeculae, is the weakest physical connection between the brain and the skull. The structure of the sheep meninges in the histological image in Figure 7 is very similar to the schematic illustration of the morphology of human meninges reported by Haines [1].

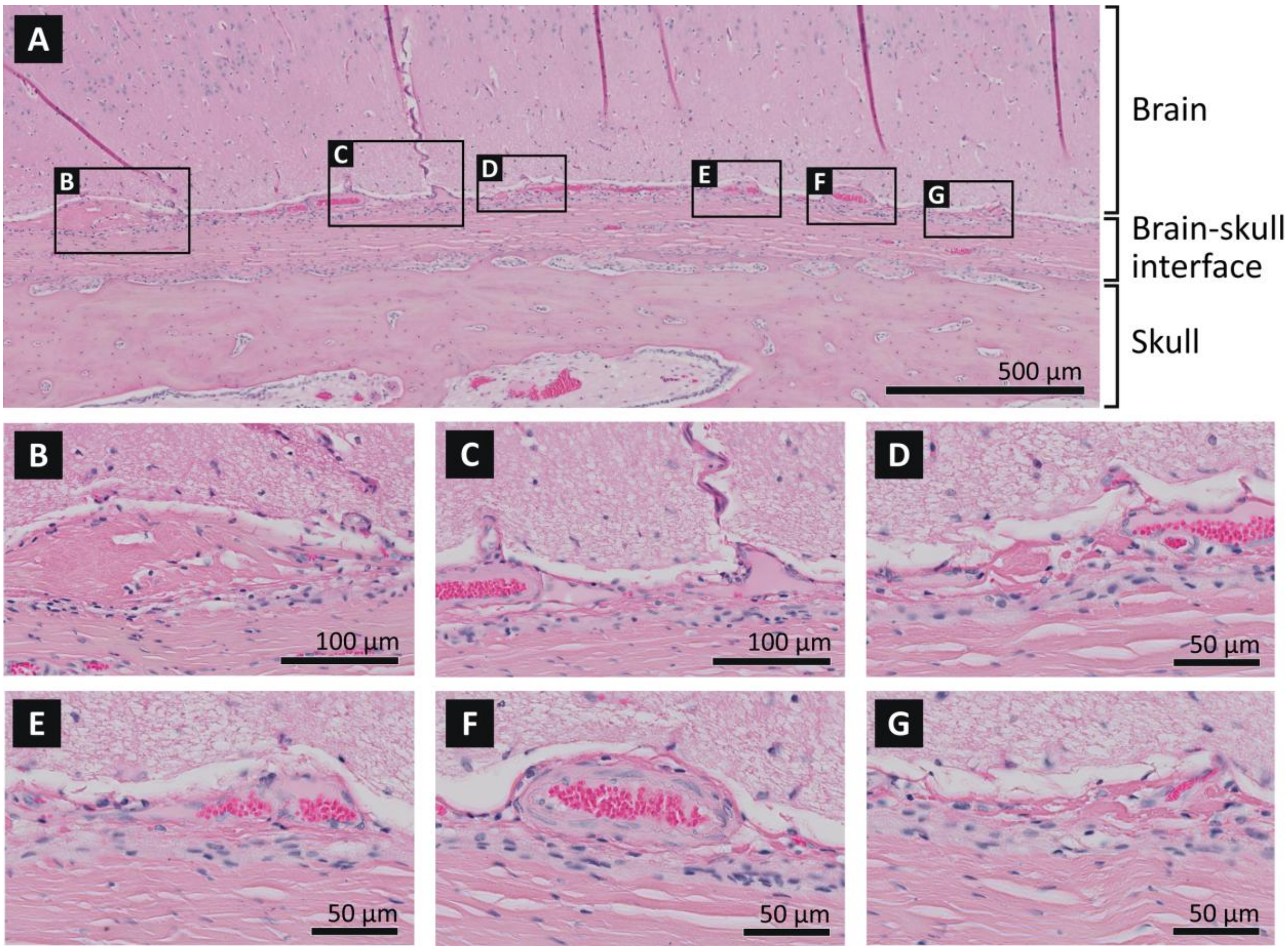

*Figure 7 Histology of the intact brain-skull interface. (A) overview of the sample with the brain tissue uppermost, the skull bone lowermost and the brain-skull interface between them. The boxes mark the subarachnoid space reproduced at higher magnification in (B) to (G). (B-G) The interface at higher magnification. The membranes are continuous along the entire length of the section and remain attached both to the cortical surface of the brain and to the inner table of the skull, with no separation at either boundary.*

### *3.4. Comparison with Established Methods for Modelling the Brain-Skull Interface*

Figure 8 compares the results obtained using a cohesive zone model of the brain-skull interface with the two approaches commonly used in biomechanical head models described in the literature [4, 9]: (i) a rigid connection between the brain cortical surface and the skull bone, and (ii) a frictionless sliding contact with separation between the brain cortical surface and skull bone. The rigid connection overestimated the force in compression, tension and shear and, since it permits no separation, cannot reproduce the failure of the interface (Figure 8a). The frictionless sliding contact slightly underestimated the compressive force and could not represent the tensile response at all since it transmits no tensile force (Figure 8a). It also underestimated the force by an order of magnitude, because without friction or adhesion only the geometric interlocking of the curved surfaces of the brain tissue and skull bone resists sliding (Figure 8a-b). Only the cohesive zone model accurately reproduced the force-displacement curves for all three loading modes together with the initiation and the progression of failure within the interface (Figure 8a-b).

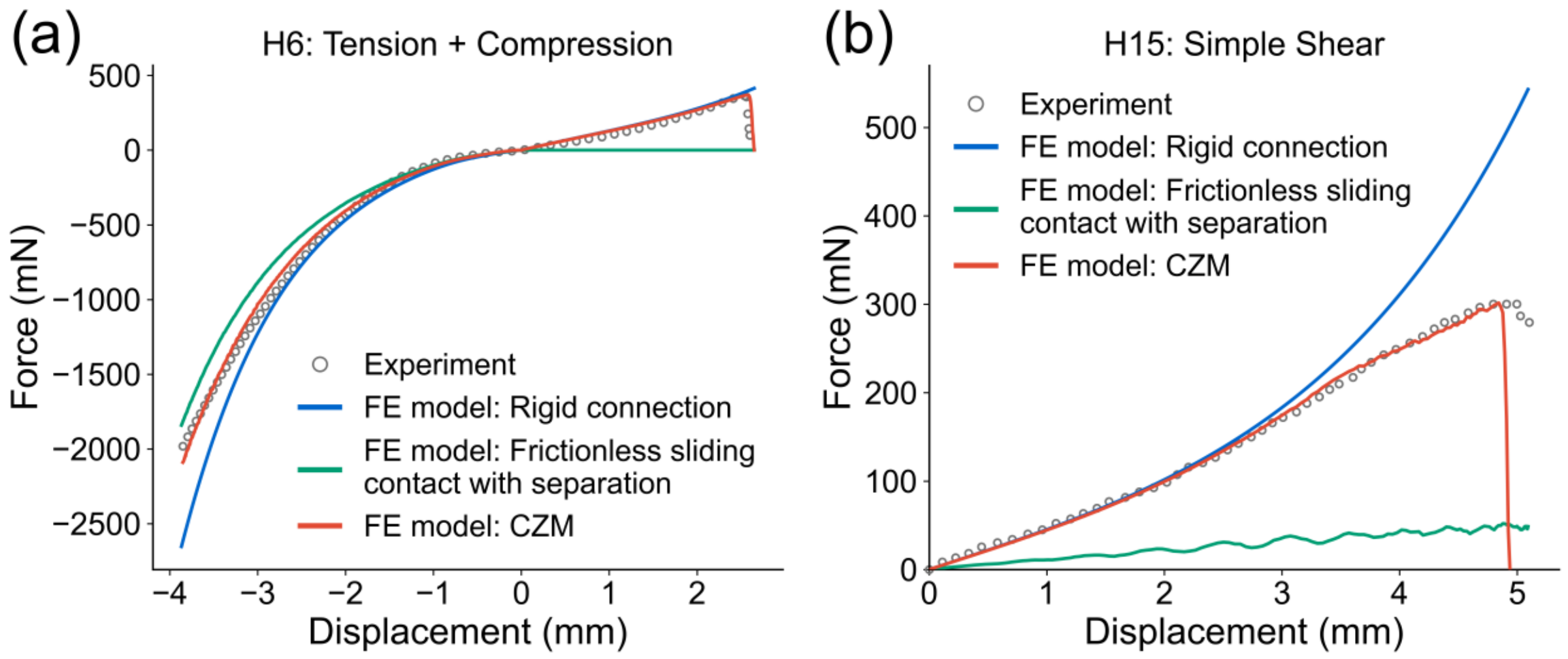

*Figure 8 Force-displacement characteristics of brain-skull complex samples obtained from the experiments (grey hollow circles) and the calibrated FE models with three different representations of the brain-skull interface, rigid connection between brain tissue and skull (Blue solid line), frictionless sliding contact with separation between brain tissue and skull (green solid lines) and a subject-specific calibrated cohesive zone model (CZM) representing the brain-skull interface (red solid lines). (a) Subject H6 (tension and compression at 0.02 $s^{-1}$). (b) Subject H15 (simple shear at 0.02 $s^{-1}$). For clarity, the experimental data are shown at reduced point density.*

## 4. Discussion

We characterised the intact ovine brain-skull interface under tension, compression and shear, and at three strain rates, within a single experimental and modelling protocol (Figure 2). We used a hyperelastic second-order Ogden model to represent the constitutive behaviour of the brain tissue and a cohesive zone model for the brain-skull interface.

The constitutive parameters of the brain tissue were first identified from the experiments using the brain tissue samples and then held constant when modelling the experiments using the brain-skull complex sample (consisting of the brain tissue, brain-skull interface, and skull bone) extracted from the same subject, so that the parameters of the cohesive zone model representing the interface were the only unknowns. The model for every experiment used the sample-specific geometry reconstructed from the sample MRI, as we have shown previously that accurate representation of the geometry is a sine qua non when determining constitutive parameters of very soft tissues [19] and that idealised geometries introduce bias in the identified parameters. Histological analysis confirmed that the meninges in the brain-skull complex samples were intact and attached to both the skull and the brain (Figure 7), demonstrating the validity of our sample preparation protocol.

Brain tissue shear moduli identified in this study (0.77 to 1.97 kPa) are consistent with those we obtained previously for ovine brain tissue at 0.02 $s^{-1}$ using the same protocol and sample-specific geometries [15, 16, 19] and lie within the range reported in the literature for brain tissue at comparable strain rates [20, 23, 24, 35, 45-47]. The increase in shear modulus with strain rate is consistent with the well-documented behaviour of the brain tissue [21, 23, 24, 27, 35, 46, 47]. The agreement between the shear moduli identified from the shear tests and from the tension and compression tests at the same rate indicates that the second-order Ogden model with fixed exponents describes the tissue adequately in all three loading modes.

The calibrated parameters of the cohesive zone model applied to model the brain-skull interface were of the same order for all loading modes: the maximum normal traction was 2.4 ± 0.8 kPa, the maximum tangential traction was 4.1 ± 1.0 kPa, and the fracture energy was 0.35 ± 0.14 N $m^{-1}$. The parameters identified from the shear tests did not significantly differ from those identified from the tension and compression tests. These results support that the traction-separation law we selected accurately reproduces both tensile and shear failure modes of the brain-skull interface, which makes it suitable for application in computational biomechanics head models representing real-world loading scenarios.

For each loading mode, the brain tissue sample and its brain-skull complex counterpart were extracted from the same region of opposite hemispheres and contained brain tissue of similar size. The brain tissue in the complex samples therefore underwent comparable forces and displacements up to interface failure as those applied to the brain tissue samples, which sustained them without failing. Our visual inspection confirmed this: in every brain-skull complex sample, the brain tissue remained intact, and the separation occurred at the interface rather than in the tissue. For the linear softening law used here, the separation at complete failure is 2G/$t_n^0$ (where G=fracture energy and $t_n^0$=maximum normal traction) in the normal direction and 2G/$t_s^0$ (where $t_s^0$=maximum tangential traction) in the tangential direction, which, for the sample-mean parameters, gives average values of approximately 0.3 mm for normal loading and 0.2 mm for shear loading. These imply that the brain-skull interface undergoes a total failure (i.e. loses all its load-carrying capacity) when the overall interface deformation is around 30% of the initial thickness of 1 mm in the normal direction and 20% in shear.

In all samples analysed here, the brain-skull interface separated within the subarachnoid space, leaving the dura mater attached to the skull and the pia mater attached to the brain. This is the most compliant part of the brain-skull interface, being orders of magnitude softer than the dura mater [13, 14]. This is consistent with the histological analysis conducted in this study (Figure 7) and reported in the literature [1], which shows that the subarachnoid space is bridged by a sparse network of trabeculae (Figure 7).

We did not find a statistically significant dependence on strain rate for the constitutive parameters of the cohesive zone model representing the brain-skull interface under tension-compression loading (exact Kruskal-Wallis test, Table 3). However, as each loading group consisted of only four or five subjects and the parameters of the brain-skull interface varied considerably between subjects (Table 2), a moderate effect of the strain rate on these parameters may not have been detected.

The failure of the cohesive zone model in the FE models of brain-skull complex samples was often sudden, whereas in some experiments the brain-skull interface failed progressively, with the force remaining close to its peak while the failure propagated across the interface. One possible reason can be that we assumed uniform interface strength which implies that most cohesive elements reach the damage initiation criterion at nearly the same time. However, as the arachnoid trabeculae density is not uniform [2], local variation in the interface strength may be anticipated. Evaluating this hypothesis would require separate investigation and, therefore, is beyond the scope of this study.

Although we calibrated the cohesive zone model parameters separately for each subject, the sample-specific FE models of the typical experiments, with the subject's own brain tissue parameters and the mean interface parameters of all 19 subjects, also accurately reproduced the measured force-displacement curves up to the onset of interface damage and predicted the failure point with differences comparable to the inter-subject variation of the interface parameters (Section 3.2, Figure 6, and Table 2). This indicates that the mean parameters provide an adequate representation of the brain-skull interface.

The comparison in Figure 8 shows that the approach used to model the brain-skull interface affects the force-displacement response of the brain-skull complex samples. The rigid connection overestimated the force in all loading modes and, since it permits no separation, cannot predict the failure of the interface (Figure 8a-b). The frictionless sliding contact with separation could not represent the tensile and shear resistance of the brain-skull interface, since it transmits only compressive force (Figure 8a-b). Only the cohesive zone model reproduced both the force transmitted by the interface before failure and the force drop after the failure. We compared the effect of brain-skull interface modelling for one sample in each loading mode; however, the differences between the three representations are large and are unlikely to depend on the sample chosen. A systematic assessment of how the proposed cohesive zone model representation of the brain-skull interface affects the predictions of computational whole-head models, compared with the conventional representations, remains to be studied.

Several limitations of this study should be noted. The samples were extracted from sheep cadaver heads. Although the overall anatomy of the sheep meninges is similar to that of the human meninges, the sheep meninges are thinner, and their strength may differ from that of the human meninges. The samples were extracted from the convexity of one hemisphere and did not include the falx cerebri and the bridging veins. Therefore, the identified parameters describe the brain-skull interface over the convexity only. The falx cerebri and the bridging veins anchor the brain to the skull far more firmly than the brain-skull interface, and they require a separate representation in computational head models. As the fastest strain-rate used in this study was 0.2 $s^{-1}$, it may be argued that our conclusions regarding the strain-rate dependency of the brain-skull interface biomechanical properties require further verification using much faster strain rates.

## 5. Conclusions

We determined the mechanical behaviour of the intact ovine brain-skull interface under tension, compression and shear by testing brain-skull complex samples excised from 19 sheep cadaver heads at strain rates of 0.002, 0.02 and 0.2 $s^{-1}$ in tension and compression and 0.02 $s^{-1}$ in shear. Histology confirmed that the meninges remained intact and attached to the brain and the skull after sample excision. We demonstrated that a cohesive zone model accurately captures the brain-skull interface responses and calibrated its parameters by inverse finite element analysis of sample-specific models, with the brain tissue parameters calibrated beforehand for each subject. The calibrated maximum normal traction was 2.35 ± 0.81 kPa, the maximum tangential traction 4.14 ± 0.97 kPa and the fracture energy 0.35 ± 0.14 N $m^{-1}$. These parameters did not differ significantly between shear and tension-compression loading, and no statistically significant dependence on strain rate was found within the analysed range, whereas the shear modulus of the brain tissue increased from 0.94 ± 0.16 kPa at 0.002 $s^{-1}$ to 1.60 ± 0.30 kPa at 0.2 $s^{-1}$. The models with the sample-specific calibrated cohesive zone model parameters reproduced the measured force-displacement curves and the onset of interface failure closely. Neither a rigid connection nor a frictionless sliding contact between the brain and the skull, the approaches commonly used in computational head models, reproduced the measured responses or the failure of the brain-skull interface.

These results provide an experimentally determined constitutive description of the intact brain-skull interface, together with the brain tissue parameters identified in the same subjects, that can replace the ad hoc boundary/contact conditions for the brain currently used in computational head models.

## Acknowledgements

The authors acknowledge the facilities and scientific and technical assistance of the Western Australia National Imaging Facility (WA NIF) and thank Dr Tim Rosenow and Dr Sjoerd B. Vos of WA NIF and The University of Western Australia (UWA) for their support with image acquisition. The authors also thank Associate Professor Elena Juan Pardo, Head of the Translational 3D Printing Laboratory for Advanced Tissue Engineering (T3mPLATE) at the Harry Perkins Institute of Medical Research (Nedlands, WA, Australia), and the T3mPLATE laboratory personnel for providing access to the experimental equipment. The authors acknowledge Associate Professor Stuart Hodgetts (Spinal Cord Repair Laboratory, School of Human Sciences, UWA) for his advice on sample dissection and storage, Mrs Mary Lee and Dr Shirley Chang, CELLCentral, School of Human Sciences, UWA, for the histological preparation, the Centre for Microscopy, Characterisation, and Analysis at The University of Western Australia, a Microscopy Australia (ROR: 042mm0k03) facility enabled by NCRIS, and Ulani Hayter Otaola, an Honours student in the School of Engineering, for her assistance with conducting the experiments.

## Funding

This research was supported by the Australian Government through the Australian Research Council's Discovery Projects funding scheme (Project DP230100949). Author S. A. acknowledges support of the University of Western Australia Scholarship for International Research Fees and the University of Western Australia - Intelligent Systems for Medicine

Laboratory (ISML) Higher Degree by Research Scholarship in Computational Biomechanics (partly funded from the Australian Research Council's ARC Discovery Project DP230100949).

## Declaration of competing interest

The authors declare that they have no known competing financial interests or personal relationships that could have appeared to influence the work reported in this paper.

## Ethical approval

All experiments were conducted under The University of Western Australia Biosafety Approval F69199 and Harry Perkins Institute of Medical Research Institutional Biosafety Committee Approval RA 23/2023. Sheep cadaver heads were an abattoir by-product (no live-animal work), and the study was therefore exempt from animal ethics approval.

## Data availability

The data that support the findings of this study will be made available upon acceptance.

## Supplementary material

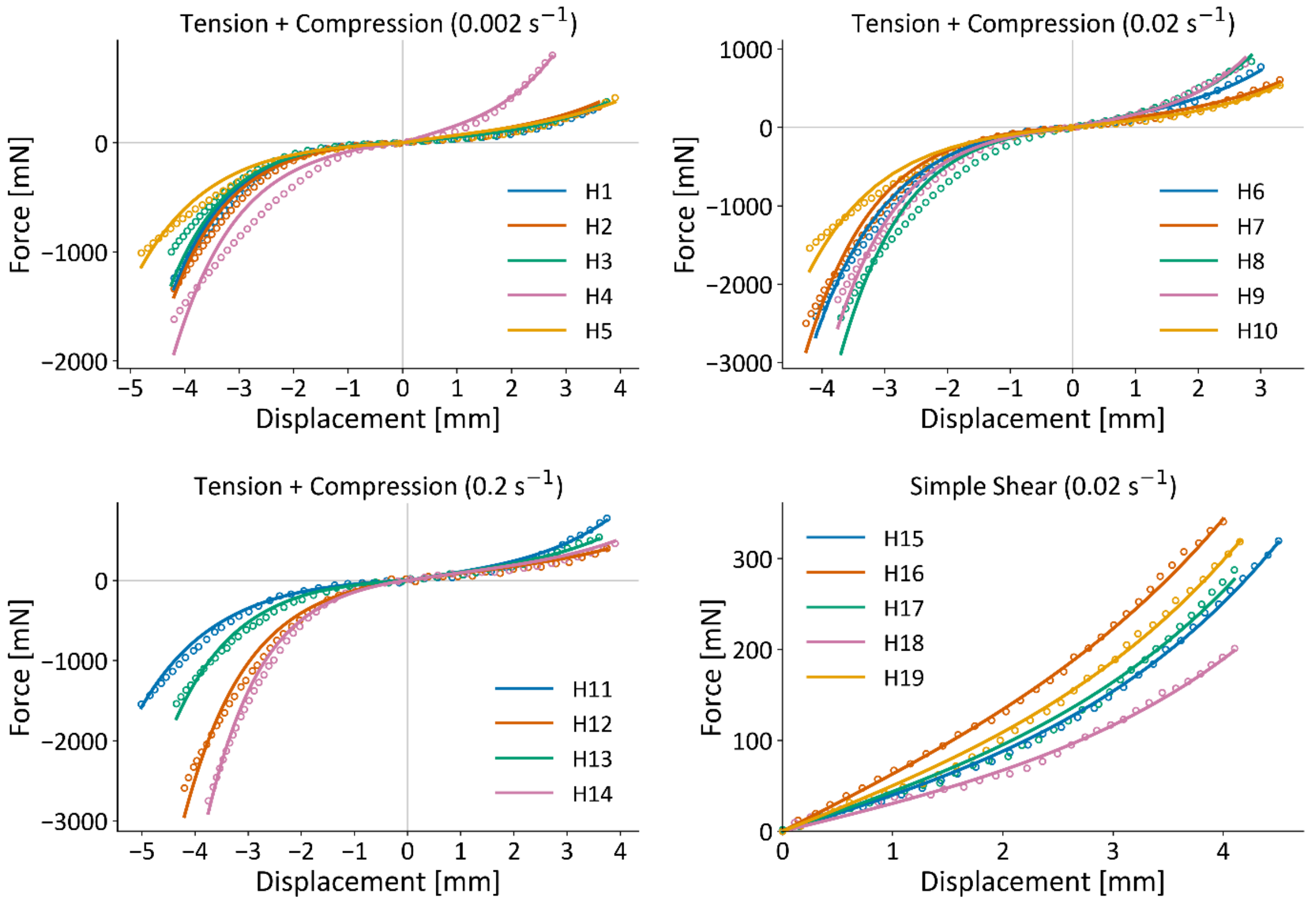

*Figure S1. Force-displacement characteristics of brain tissue samples, each from a different subject (Table 2): experiments (grey hollow circles) and sample-specific FE model of the same experiment (red solid line), with the shear modulus of the brain tissue calibrated for that subject. (a) tension and compression tests at a strain rate of 0.002 $s^{-1}$; (b) tension and compression tests at a strain rate of 0.02 $s^{-1}$; (c) tension and compression tests at a strain rate of 0.2 $s^{-1}$; (d) simple shear test at a shear strain rate of 0.02 $s^{-1}$. Negative displacement and force denote compression.*

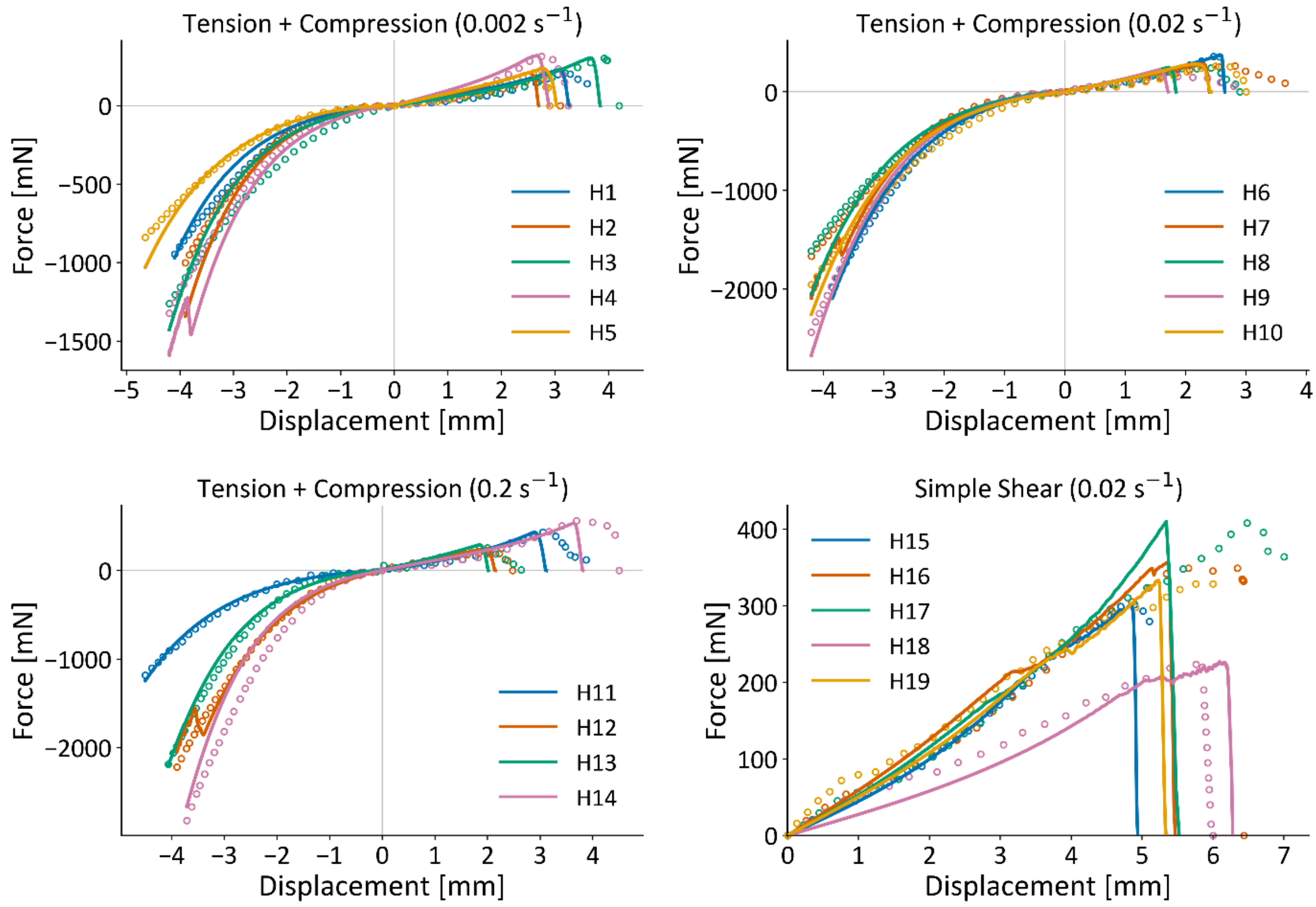

*Figure S2. Force-displacement characteristics of brain-skull complex samples from the same subjects as in Figure S1: experimental results (grey hollow circles) and the sample-specific FE model of the same experiment, with the brain-skull interface represented by a cohesive zone model (CZM) with the brain-skull interface parameters ($t_n^0$, $t_s^0$ and G) calibrated for that subject (red solid lines) or the mean interface parameters of all 19 subjects (blue dashed lines, Table 2). (a) tension and compression tests at a strain rate of 0.002 s$^{-1}$; (b) tension and compression tests at a strain rate of 0.02 s$^{-1}$; (c) tension and compression tests at a strain rate of 0.2 s$^{-1}$; (d) simple shear test at a shear strain rate of 0.02 s$^{-1}$. Negative displacement and force denote compression. The drop in force in tension and in shear marks failure of the brain-skull interface.*